\documentclass {aa}
\usepackage{times}
\usepackage{epsfig}
\usepackage{latexsym}
\usepackage{mymacros}
\usepackage{natbib}                        
\newcommand{\gVelorum}{$\gamma^2$\,Velorum}
\newcommand{\gVel}{$\gamma^2$\,Vel}
\newcommand{\Mi}{\ensuremath{M_\mathrm{i}}}
\newcommand{\D}{\displaystyle}
\newlength{\mylen}
\newlength{\colone}
\newlength{\coltwo}
\newlength{\colwidth}
\newlength{\fignarrow}
\newlength{\figwide}
\graphicspath{{figures/}}
\begin{document}
\thesaurus{1                            
           (13.07.2;                    
           02.14.1;             
           08.09.2 \gVel;               
           08.23.2;                     
           09.09.1 IRAS Vela shell)}    
\title{\boldmath%
        COMPTEL limits on \Al26\ 1.809~\MeV\ line emission from \gVelorum} 

\author{U.~Oberlack\inst{1,6} \and U.~Wessolowski\inst{1} \and R.~Diehl\inst{1} 
        \and K.~Bennett\inst{4} \and H.~Bloemen\inst{2} \and W.~Hermsen\inst{2} 
        \and J.~Kn\"odlseder\inst{5} \and D.~Morris\inst{3}
        \and V.~Sch\"onfelder\inst{1} \and P.~von~Ballmoos\inst{5} 
        }

\offprints{Uwe Oberlack}
\mail{oberlack@astro.columbia.edu}

\institute{Max-Planck-Institut f\"ur extraterrestrische Physik, 
           D-85740 Garching, Germany 
           \and 
           SRON Utrecht, NL-3584 CA Utrecht, The Netherlands
           \and 
           Space Science Center, University of New Hampshire, 
           Durham NH 03824, USA 
           \and Astrophysics Division, ESTEC, 
           NL-2200 AG Noordwijk, The Netherlands 
           \and
           Centre d'Etude Spatiale des Rayonnements (CNRS/UPS), 
           F-31028 Toulouse, France
           \and
           Columbia Astrophysics Laboratory, Columbia University, 
           New York NY 10027, USA}

\date{Received \ldots; accepted October 28, 1999}

\titlerunning{COMPTEL Limits on \Al26\ 1.809~\MeV\ line emission from \gVel}
\authorrunning{U. Oberlack et al.}
\maketitle 

\begin{abstract}
The Wolf-Rayet binary system \object{\gVel} (\object{WR\,11}) is the closest
known Wolf-Rayet (WR) star. Recently, its distance has been redetermined by
parallax measurements with the HIPPARCOS astrometric satellite yielding
$258^{+41}_{-31}$~pc, significantly lower than previous estimates (300 --
450~pc). Wolf-Rayet stars have been proposed as a major source of the Galactic
\Al26\ observed at 1.809~\MeV. The gamma-ray telescope COMPTEL has previously
reported 1.8~\MeV\ emission from the Vela region, yet located closer to the
Galactic plane than the position of \gVel. We derive an upper 1.8~\MeV\ flux
limit of $1.1 \; 10^{-5}$~\flux\ ($2~\sigma$) for the WR star. With the new
distance estimate, COMPTEL measurements place a limit of
$\left(6.3^{+2.1}_{-1.4}\right) \; 10^{-5}$~\Msol on the \Al26\ yield of \gVel,
thus constrains theories of nucleosynthesis in Wolf-Rayet stars. We discuss the
implications in the context of the binary nature of \gVel\ and present a new
interpretation of the IRAS Vela shell.
\end{abstract}

\section{Introduction}
The 1.809 \MeV\ gamma-ray line from radioactive decay of \Al26\ 
$[$mean lifetime $(1.07\pm 0.04) \; 10^6$~yr \cite[]{PEndt:90}$]$ traces
recent nucleosynthesis in the Galaxy. This was the first gamma-ray line detected
from the interstellar medium \cite[]{WMahoney:84}, and had been predicted from
nucleosynthesis calculations of explosive carbon burning in core-collapse 
supernovae \cite[]{RRamaty:77,WArnett:77}. Other \Al26\ production
sites have been proposed as well, covering a wide range of densities (0.5 --
$3\;10^5$~\dens), temperatures ($3\;10^7$ -- $3\;10^9$~K), and time scales (1 --
$10^{14}$~s) at which proton capture on \nuc{25}{Mg} (within the Mg-Al chain)
would create \Al26: explosive hydrogen burning in novae and supernovae
\cite[]{MArnould:80:al26:h_burning}, neon burning in the presupernova and
supernova stage \cite[e.g.][]{SWoosley:95:TWeaver}, neutrino-induced
nucleosynthesis in supernovae \cite[]{SWoosley:90:nu_process}, convective core
hydrogen burning in very massive stars \cite[]{DDearborn:85}, and hydrogen shell
burning or ``hot bottom burning'' at the base of the convective stellar envelope
in asymptotic giant branch (AGB) stars \cite[]{HNorgaard:80:al26_agb,
MForestini:91:al26_agb}. \Al26\ nucleosynthesis and observations have been
reviewed by \cite{DClayton:87:MLeising,GMacPherson:95:solar_system,
NPrantzos:96:RDiehl,RDiehl:98:FTimmes:PASP}.

Current theories still predict significant amounts of \Al26\ from core-collapse
supernovae \cite[]{SWoosley:95:TWeaver, FThielemann:96:collaps_sne,FTimmes:95:al26_fe60}, the wind 
phases of very massive stars where \Al26\ is produced on the main sequence and
ejected mainly in the Wolf-Rayet stage \cite[]{NLanger:95:al26,GMeynet:97:al26}
or possibly (in case of fast rotating stars) in a red supergiant stage as well
\cite[]{NLanger:97:rotation}, and from the most massive AGB stars
\cite[]{GBazan:93:al26_agb}. The expected \Al26\ contribution of chemically
enriched novae lowered after major revisions of key reaction rates
\cite[]{JJose:97:al26_novae,SStarrfield:98:nova_models}. All models
suffer from large uncertainties in \Al26\ yields, ranging from factors of three to
orders of magnitude for the various proposed astrophysical sites.

 Recent results from the  COMPTEL telescope
aboard the Compton Gamma-Ray Observatory showed growing evidence for a
young (massive) population dominating the galaxy-wide \Al26\ production
\cite[]{RDiehl:95:galactic,JKnoedlseder:96:spiralmodelle,UOberlack:96:cgro95,
UOberlack:97:thesis,JKnoedlseder:97:thesis}. While low-mass AGB stars and novae
can be ruled out as the main \Al26\ source, a distinction between supernovae
and hydrostatic production in very massive stars appears difficult due to the
similar evolutionary time scales involved.%
\footnote{See \cite{JKnoedlseder:99:al26_origin} for arguments favouring WR
          stars as the dominant source of Galactic \Al26.}
Therefore, detection of individual objects would be essential as calibrator,
but the sensitivity of current instruments restricts this approach to very few
objects. Upper limits for five individual supernova remnants have been derived
by \cite{JKnoedlseder:96:snr_uplim} and the possibility of interpreting
1.8~\MeV\ emission from the Vela region with individual objects has been
discussed by \cite{UOberlack:94:vela} and \cite{RDiehl:99:vela}.

\gVel\ \cite[WR\,11,][]{KvanderHucht:81:wr_catalogue} is a WC8$+$O8--8.5III 
binary system at $(l,b) = (262.8\deg,-7.7\deg)$, containing the
nearest Wolf-Rayet star to the Sun at a distance of $258^{+41}_{-31}$~pc, as
determined by parallax measurements with the HIPPARCOS satellite
\cite[]{KvanderHucht:97:wr11,DSchaerer:97:wr11}. $[$Another recent investigation
determines a spectral type of O7.5 for the O star
\cite[]{ODeMarco:99:gamma_vel}.$]$
A previous 1.8~\MeV\ flux limit(2~$\sigma$) for \gVel\ of $1.9\;10^{-5}$~\flux\
had been determined by \cite{RDiehl:95:vela}, based on observations of the
first 2~1/2 CGRO mission years.

In Sect.~2, we describe our search for 1.8~\MeV\ emission from \gVel\ and the
derivation of upper flux limits for several emission models. Sect.~3 discusses 
the initial mass range of the WR star, implications of our flux limit for
stellar models, and proposes an alternative interpretation of the ``IRAS Vela
shell''. We summarize in Sect.~4.

\section{Data analysis and source models} 
The COMPTEL telescope spans an energy range from 0.75 to 30~\MeV\ with spectral
resolution of 8\% (FWHM) at 1.8~\MeV, and performs imaging with an angular
resolution of 3.8\deg\ FWHM at 1.8~\MeV\ within a \mbox{$\sim 1$ sr} field of
view. It features the Compton scattering detection principle through a
coincidence measurement in two detector planes (see \cite{VSchoenfelder:93} for
details). Imaging analysis and model fitting occurs in a three-dimensional
dataspace consisting of angles $(\chi,\psi,\phibar)$ describing the scattered
photon's direction and the estimated Compton scatter angle, respectively
\cite[]{RDiehl:92:dataanalysis4}.

In this paper, we concentrate on fitting sky models, convolved with the
instrumental response, in the imaging dataspace. The present analysis makes use
of all
data from observations 0.1 -- 522.5 which have been combined in a full-sky
dataset, comprising 5 years of observing time between May 1991 and June 1996.
Events were collected within a 200~\keV\ wide energy window from (1.7 --
1.9)~\MeV\ into the imaging dataspace with $1\deg \times 1\deg$ binning in
$(\chi,\psi)$ (in galactic coordinates) and 2\deg\ binning in \phibar. The
dataspace has been restricted to $l \ininterval{185\deg,320\deg}$ and $|b| \le
50\deg$ to concentrate on emission from the Vela region, but not to loose
information from the up to 100\deg\ wide response cone (at $\phibar =
50\deg$). The instrumental and celestial continuum background have been modelled
by interpolation from adjacent energies, with enhancements from Monte Carlo
modelling of identified activation background components. A previous version of
the background handling and event selections have been described by
\cite{UOberlack:96:cgro95} and \cite{JKnoedlseder:96:SPIE}, more details on the
recently improved background handling and the complete dataset are reported in
\cite{UOberlack:97:thesis} [Oberlack et al., in prep.]. 

For the derivation of upper limits, the maximum likelihood ratio test has been
applied \cite[]{WCash:79}: A null hypothesis $H_0$ is compared with an
extended alternative hypothesis $H_1$, which includes $q$ additional continuous
model parameters, using the likelihood function $\cal L$,
which is the product of the Poisson probabilities $p_k$ in $N$ dataspace
cells:
\begin{equation}
   {\cal L} = \prod_{k=1}^{N} p_k \qquad 
   p_k = \left \{ 
   \begin{array}{ll}
   \D \frac{\mu_k^{n_k}}{n_k!} \: e^{-\mu_k} & \textrm{ for } \; \mu_k > 0 \\
      1 \quad  & \textrm{ for } \; \mu_k = 0 , \; n_k = 0  \\
      0 \quad  & \textrm{ for } \; \mu_k = 0 , \; n_k > 0 \\
   \end{array}
   \right. 
\end{equation}
where $n_k$ is the number of counts in cell $k$ and
\begin{equation}
   \mu_k = \sum_{s=1}^{n_\mathrm{src}} a^{(s)} \: \mu_k^{(s)} 
                                     + b \: \mu_k^\mathrm{bgd} 
\end{equation}
is the predicted number of counts due to sources and the background model, which
includes the scaling parameters $a^{(s)}$, $b$ varied in the fit. Each source
model $(s)$ can be described by a (normalized) flux map $\{f_j^{(s)}\}$
convolved with the response matrix $R_{jk}$:
\begin{equation}
   \mu_k^{(s)} = \sum_{j} R_{jk} f_j^{(s)}
\end{equation}

If the null hypothesis is true, the probability distribution of the ratio of the
maximum likelihood $\hat{\cal L}_1$ achieved by fitting $H_1$ to the data over
the maximum likelihood $\hat{\cal L}_0$ achieved by fitting $H_0$
to the data can be described analytically:
\begin{equation}
p\left(2 \; \ln\left( \frac{\hat{\cal L}_1}{\hat{\cal L}_0}\right)\right) 
= p(\chi^2_q)
\end{equation}
where $p(\chi^2_q)$ is the tabulated $\chi^2$ probability distribution with $q$
degrees of freedom. 

\begin{figure}
\centering
\psfig{file=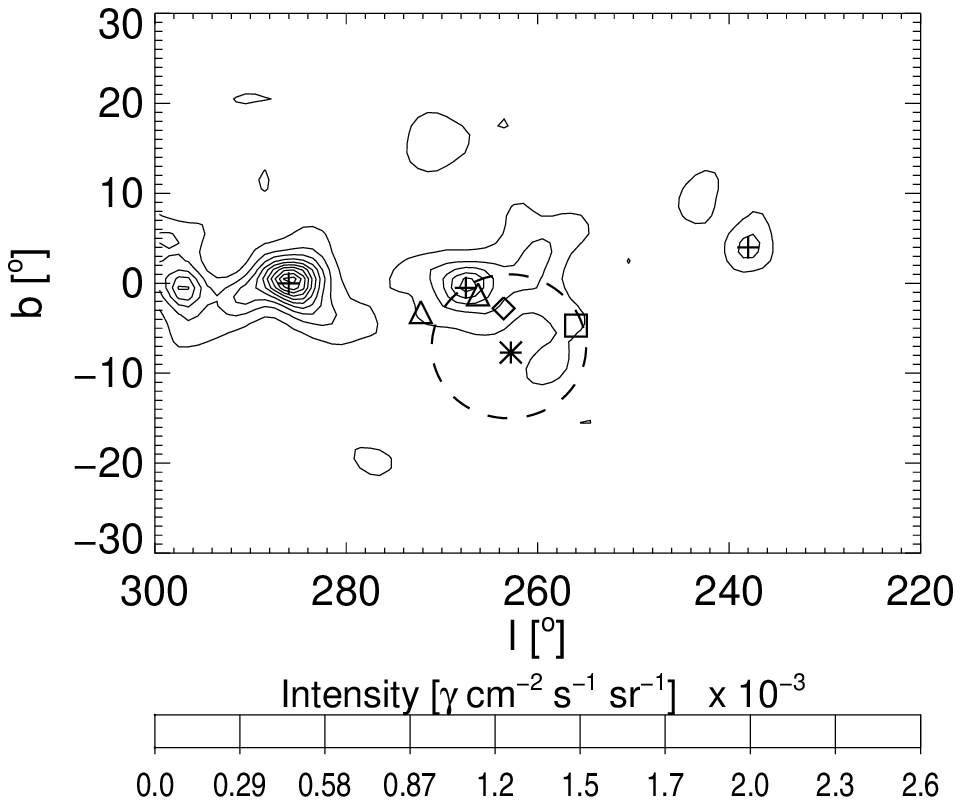,width=\fignarrow,%
       bbllx=4,bblly=43,bburx=283,bbury=235,clip=}
\caption[]{COMPTEL 1.8~\MeV\ map of the Carina / Vela / Puppis regions based on
a deconvolution using the Maximum Entropy Method \cite[]{UOberlack:97:thesis}.
Contour lines are equally spaced with steps of $2.9 \: 10^{-4}$~\intens. The
dashed line indicates the IRAS Vela shell, the marked point source models are
listed in Table~\ref{t:model_b}.} 
\label{f:vela-map} 
\end{figure}

Fig.~\ref{f:vela-map} shows a Maximum-Entropy deconvolved map of the 1.8~\MeV\
emission from the Carina / Vela / Puppis regions. While no 1.8~\MeV\ flux is
detected from the position of \gVel, significant extended emission is observed
in nearby regions, with an intensity peak around $(267\fdg5,-0\fdg5)$. Due to
the broad COMPTEL response, such emission needs to be modelled. We test its
impact on flux limits for \gVel\ with four different emission models (additional
to the background model), guided by the deconvolved image, by candidate \Al26\
sources in the region, and by results on the galaxy-wide \Al26\
distribution. These are the tested source models in addition to the background
model:
\begin{enumerate}
\renewcommand{\labelenumi}{\alph{enumi}.}
\parsep 0pt \itemsep 0pt
\item A single point source at the position of \gVel. 
\item A detailed model describing the observed emission empirically and
      including 1.8~\MeV\ candidate sources of the region 
      (Fig.~\ref{f:vela-map}): an exponential disk with emissivity $\propto
      \exp\{-R/R_0\} \cdot \exp\{-|z|/z_0\}$, galactocentric scale length
      $R_0=4$~kpc, and scale height $z_0=180$~pc as an approximation to the
      large-scale \Al26\ distribution, an additional homogenous stripe around 
      $l = 310\deg$ for simplified modeling of excess emission in this region
      and 8 point sources including \gVel. Details are listed in
      Table~\ref{t:model_b}.
\item An intermediate model: 2 point sources at the positions of maximum 
      intensity in Fig.~\ref{f:vela-map} at $(l,b)$: $(286\fdg0,0\fdg0)$, 
      $(267\fdg5,-0\fdg5)$ plus a point source for \gVel.
\item Like model (c) but replacing the point source for \gVel\ by a model for
      the IRAS Vela Shell: a spherical shell with a thickness of 10\% of the
      8\deg\ radius around $(l,b) = (263\deg,-7\deg)$ 
      \cite[]{MSahu:93:gum_nebula:asp}.
\end{enumerate}
\begin{table}
\setlength{\mylen}{4.4\baselineskip}
\setlength{\colone}{0.5\linewidth}
\setlength{\coltwo}{0.75\linewidth}
\tabcolsep 3.0pt                                        
\addtolength{\colone}{-2\tabcolsep}
\addtolength{\coltwo}{-2\tabcolsep}
\renewcommand{\arraystretch}{1.0} \small \centering
\vspace*{-0.5\baselineskip}
\begin{tabular*}{\linewidth}[t]{|l@{ }l@{\extracolsep{\fill}}l@{}l|} \hline
\multicolumn{4}{|c|}{\parbox{\coltwo}{Exponential disk: \hfill 
                                      $R_0=4$~kpc, $z_0=180$}} \\ \hline
\multicolumn{4}{|c|}{\parbox{\coltwo}{Homogenous stripe: \hfill 
            $300\deg \le l \le 320\deg$ \enspace $|b| \le 3\deg$}} \\ \hline
\multicolumn{4}{|c|}{Point sources $(l,b)$} \\
$\ast$ (262\fdg8,$-$7\fdg7) & \gVel &
$\scriptstyle +$ (238\fdg0,   4\fdg0) & Puppis feature \\
$\scriptstyle \Box$ (256\fdg0,$-$4\fdg7) & $\zeta$~Pup &
$\scriptstyle \Diamond$ (263\fdg6,$-$2\fdg8) & Vela pulsar \\
$\scriptstyle \triangle$ (266\fdg2,$-$1\fdg2) & SNR ``Vela Junior''$^\S$ &
$\scriptstyle +$ (267\fdg5,$-$0\fdg5) & Vela feature   \\
$\scriptstyle \triangle$ (272\fdg2,$-$3\fdg2) & SNR G$272.2-3.3$ &
$\scriptstyle +$ (286\fdg0,   0\fdg0) & Carina feature \\ \hline
\end{tabular*}
\caption[]{Source components of the detailed model (b). For the exponential
disk, scale length and scale height are given as well as the range of the
homogenous stripe. Point sources reflect candidate \Al26\ sources in the region
and additional emission features appearing in the map. 
$^\S$\cite{BAschenbach:98:vela_junior,AIyudin:98:44Ti_Vela}}
\label{t:model_b}
\vspace*{-2\baselineskip}
\end{table}

The first three models assume that all \Al26\ from \gVel\ is kept within an
observed 'ejecta-type' wind shell around the binary system with a diameter of
$57'$ \cite[]{AMarston:94:survey2}. Given COMPTEL's angular resolution, this can
be modelled by a point source. For completeness, we consider a fourth model,
which places \Al26\ into the so-called IRAS Vela shell, an extended ($\sim
8\deg$ radius) structure almost centered on \gVel, which has been identified by
\cite{MSahu:92:thesis} from IRAS 25$\mu$m --
100$\mu$m maps and investigations of cometary globules. This is a region of
relatively bright \Halpha\ emission within the \object{Gum nebula}
\cite[]{AChanot:83:gum_nebula}, where young stellar objects are forming
\cite[]{TPrusti:92:YSOs}.  \citeauthor{MSahu:92:thesis} found the Vela shell to
be a structure distinct from the Gum nebula and presented an interpretation as
supershell from the aged Vela~OB2 association of which \gVel\ may be a member
\cite[]{PdeZeeuw:97:OB_ass}. (As a consequence of HIPPARCOS parallaxes, \gVel\
would be located on the very near side of the association.)  In this case, part
of the ejected \Al26\ may have traversed the gas bubble and be accumulated in
the dense outer shell. We discuss a different interpretation, however, in
Section~\ref{ss:iras_shell}.

\begin{figure}
\centering
\psfig{file=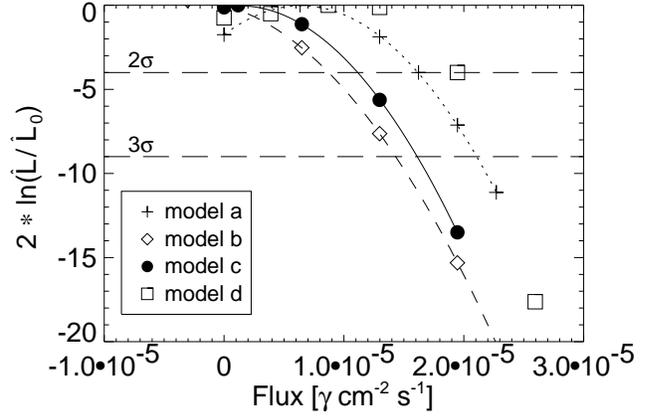,width=\fignarrow}
\caption[]{Determination of upper flux limits with the maximum likelihood ratio
test using 4 different emission models (see text).} 
\label{f:uplim} 
\vspace*{-0.5\baselineskip}
\end{figure}  
Fig.~\ref{f:uplim} shows the logarithmic maximum likelihood ratio
$2\,\ln(\hat{\mathcal{L}} / \hat{\mathcal{L}_0})$ as a function of model flux.
No significant flux from \gVel\ is found for any of the tested models. The
following 2$\,\sigma$ upper limits are derived:
\begin{displaymath}
\begin{array}[t]{lrcl}
  \textrm{Model a:} & f &<& 1.6 \; 10^{-5}~\flux \\
  \textrm{Model b:} & f &<& 0.9 \; 10^{-5}~\flux \\
  \textrm{Model c:} & f &<& 1.1 \; 10^{-5}~\flux \\
  \textrm{Model d:} & f &<& 1.9 \; 10^{-5}~\flux \\
\end{array}
\end{displaymath}
Given our 1.8 MeV map of the region, model~(a) seems overly simplistic. Lacking
alternative source models, part of the observed Vela emission is attributed to
\gVel\ and yields the most conservative upper limit for a point source
model. Model~(b) contains many free parameters and may approach an over-fit of
the data, where statistical fluctuations of the background are fitted. Although
the fitted background scaling factor is lowest for this model, the lowest (even
slightly negative) flux is attributed to \gVel. Consistent with the map in
Fig.~\ref{f:vela-map}, the flux from Vela-Puppis is better described by sources
within the galactic plane than by \gVel. Model~(c) is ``intermediate'' in that
it accounts for the strongest features in the map with a minimum of free model
parameters. We adopt its result for the \gVel\ flux limit and consider 
models~(a) and (b) as the extreme values for the systematic uncertainty due to
the choice in modelling of other emission close ($\sim 5\deg$ -- $10\deg$) to
\gVel. The limit from model~(d) is highest due to the large extent of this
source model. Yet, we do not consider this model a likely representation of
\Al26\ from \gVel, even if the structure itself may well be related to the
binary system, as discussed in the next section.

The 1.8~\MeV\ flux directly translates into the ``alive'' \Al26\
mass in the circumstellar medium, for a point source via:
\begin{displaymath}
  f_{1.8\,\mathrm{MeV}} = 1.8 \; 10^{-5} \flux\  
            \cdot \left(\frac{258}{d~[\mathrm{pc}]}\right)^2 
            \cdot \frac{M_{26}~[\Msol]}{1.0 \; 10^{-4}}
\end{displaymath}
Therefore, our $2\,\sigma$ upper flux limit 
corresponds to a maximum \Al26\ mass from \gVel\ of
\begin{equation}\label{e:gammavel:m_al}
  M_{26}^\mathrm{WR11} < \left(6.3^{+2.1}_{-1.4}\right) \; 10^{-5}~\Msol
\end{equation}
where the $1\sigma$ distance uncertainties from the HIPPARCOS measurement have
been taken into account.

\section{Discussion}
\subsection{Interpretation of the IRAS Vela shell}
\label{ss:iras_shell}
Is our upper limit for the \Al26\ yield from \gVel\ realistic, being derived for
a point source model, or should we rather consider the higher value implied by
potential \Al26\ accumulation in the IRAS Vela shell?

The refined distance of \gVel\ suggests a new interpretation of the IRAS Vela
shell as the \emph{main sequence bubble} of the WR progenitor star in the \gVel\
system. This would make a significant \Al26\ contamination of the shell unlikely
since this isotope is expected to appear at the stellar surface only in later
stages of stellar evolution, together with the products of core hydrogen
burning, after the hydrogen shell has been expelled.  Evidence for our
interpretation stems from observations of interstellar reddening by
\cite{GFranco:90:reddening}, who studied two regions within the Gum nebula, one
of which in projection to the IRAS shell. Only this field showed clear evidence
for a dust wall at a distance of $200 \pm 20$~pc, interpreted by the authors as
the near edge of the Gum nebula. If this is now interpreted as the near edge of
the IRAS shell, its angular extent would correspond to a radius of 32~pc and the
centre would be placed at a distance of $230 \pm 20$~pc, well within the
$1\sigma$ uncertainty of the \gVel\ distance. This would argue against a supershell
interpretation since the distance to the centre of Vela~OB2, the assumed origin
of the supershell, has been precisely measured by HIPPARCOS to $415 \pm 10$~pc
\cite[]{PdeZeeuw:97:OB_ass}. Scaling down the (uncertain) mass estimate for the
shell by \citeauthor{MSahu:92:thesis} from the distance of 450~pc she assumed to
the reduced distance of \gVel, yields a mass of $\sim 2\times
10^5$~\Msol. Combined with the observed expansion velocity of $\sim 10$~\kms\
\cite[]{MSahu:93:kinematics} this corresponds to a kinetic energy of $\sim 2
\times 10^{43}$~J ($2 \times 10^{50}$~erg), within the range of stellar wind
energy release by a massive star. In a hydrodynamic model coupled with a stellar
evolution code for a 60~\Msol\ star, \cite{GGarcia:96:CSM:LBV} even find a total
energy release of $3.3 \times 10^{44}$~J (70\% - 80\% thereof ejected before a
``luminous blue variable'' [LBV] phase) with a 45~pc radius O-star bubble. This
number only scales weakly with ambient density ($\propto n^{-1/5}$, $n =
20~\textrm{cm}^{-3}$ assumed in the model), but is considered an upper limit by
the authors because effects like heat conduction or cloud evaporation are
ignored here. Within the uncertainties of the model, our interpretation of the
IRAS Vela shell seems therefore plausible. For the further discussion we will
hence adopt our 1.8~\MeV\ mass limit which was derived from the assumption that
all \Al26\ from \gVel\ is contained within a region of $\sim 1\deg$ in diameter
around \gVel.

\subsection{Current mass estimate for \gVel}
Model predictions of \Al26\ yields for massive stars are a strong function of
initial stellar mass, e.g., $M_{26} \propto \Mi^{2.8}$ \cite[]{GMeynet:97:al26}.
A direct determination of the WR star initial mass (and thus its predicted
\Al26\ yield) from comparison of luminosity and effective temperature with
stellar evolution tracks, as is typically done for other types of stars, is not
feasible due to the lack of sufficiently accurate models of WR wind atmospheres
(and the additional complications from the binary nature of this stellar
system).  While binarity makes modeling more complicated due to additional
degrees of freedom in parameter space it allows measurement of current masses,
which can be matched with theoretical predictions together with the generic
spectral type of the WR star. Spectroscopic determination of $M_{1,2} \sin^3 i$
(where $i$ is the inclination) based on Doppler-shifted absorption (O star) and
emission lines (WR star) led to contradictory results 
\cite[]{CPike:83:wr11,AMoffat:86:wr11}. A recent redetermination of orbital
parameters by \cite{WSchmutz:97:gamma_vel:orbit} yields spectroscopic masses of:
\[
M_\mathrm{WR} \sin^3 i = 6.8 \pm 0.6 \,\Msol\ \qquad 
M_\mathrm{O} \sin^3 i = 21.6 \pm 1.1 \,\Msol
\]
They reject an earlier inclination measurement from polarisation data of $i =
70\deg \pm 10\deg$ by \cite{NStLouis:88} and rather state wider inclination
limits of $57\deg < i < 86.3\deg$ from other evidence, corresponding to a factor
$1 / \sin^3 i = 1.0$ -- 1.7 or a range from 6 to 12~\Msol\ for the current mass
of the WR star. Relying on a mass-luminosity relation for the O star
from \emph{single} star evolution models, \cite{DSchaerer:97:wr11} derive
$M_\mathrm{O} = 29 \pm 4$~\Msol, which, in turn, yields a consistent, but
model-dependent, inclination estimate of $i = 65\deg \pm 8\deg$ or a mass
estimate for the WR star of $M_\mathrm{WR} = 9^{+2.5}_{-1.2}$~\Msol\
\cite[]{WSchmutz:97:gamma_vel:orbit}. A different analysis of the same spectral
data yields a slightly higher luminosity for the O star, leading to a slightly
higher O star mass estimate of $30 \pm 2$~\Msol\ \cite[]{ODeMarco:99:gamma_vel},
yet, with the same mass-luminosity relation from single-star models.

Another observational hint on the total mass of the binary system has been
derived from interferometric measurements of the major half-axis of the binary
system $a'' = (4.3 \pm 0.5)$~mas by \cite{RHanburyBrown:70}. Kepler's law
leads to a consistent, but barely constraining, mass estimate:
\begin{equation}\label{e:gammavel:mtot:formel}
   M_\mathrm{WR} + M_\mathrm{O} = \frac{(2\pi)^2}{G} \frac{(a'' \cdot d)^3}{T^2}
                                = (30 \pm 16)~\Msol
\end{equation}
The large relative uncertainty in total mass of $\sim 54\%$ is due to the
uncertainties in $a''$ and $d$ to about equal amounts. Additional systematic
errors, however, may affect the determination of $a''$ by a fit in which other
quantities like orbital parameters, inclination, and brightness ratio of the two
stars had to be taken as fixed parameters due to limited data quality. Those
values were taken from other measurements available at the time. Most notably,
an eccentricity of 0.17 had been assumed which is distinctly lower than a
recent value of $0.326 \pm 0.01$ \cite[]{WSchmutz:97:gamma_vel:orbit}. Future
interferometric measurements could improve this situation significantly.

\subsection{Single star models}
Given the remaining uncertainties in current mass determinations, not
surprisingly \emph{initial} mass estimates and predicted \Al26\ yields vary 
significantly. To start with the simpler case, we first discuss our results in
the context of single star models, assuming that the stellar structure of the
primary (defined in terms of which star evolved first, i.e.\ now the WR star)
has not been significantly altered due to the presence of its binary companion
and that all of the mass expelled from the primary has reached the ISM without
being captured by the secondary. This corresponds to fully non-conservative mass
transfer, which is usually parametrized by the fraction $\beta$ of mass that is
accreted by the secondary out of the mass lost by the primary (i.e.\ fully
non-conservative mass transfer means $\beta = 0$).  For the Geneva single star
models, a minimum initial mass of about 40~\Msol\ is required at solar
metallicities to yield a WC-type Wolf-Rayet star \cite[]{GMeynet:94:wr_models}
as observed in \gVel. Therefore, these models would predict a minimum \Al26\
yield of $5.5 \times 10^{-5}$~\Msol\ for \gVel\ with initial mass of the WR
component $M_\mathrm{i} = 40$~\Msol\ and even $1.2 \times 10^{-4}$~\Msol\ for
$M_\mathrm{i} = 60$~\Msol, the value used by \cite{GMeynet:97:al26} for the
description of \gVel. These yields are consistent with the yields predicted by
the single-star models of \cite{NLanger:95:al26}. \cite{DSchaerer:97:wr11}
estimate an initial mass of $M_\mathrm{i} = 57 \pm 15$~\Msol\ for the WR star,
using the Geneva models.

Has all \Al26\ been expelled yet or is some fraction still buried invisibly
inside the star? The spectral type of the Wolf-Rayet star is WC, i.e.\ the
stellar wind is carbon rich, which means that products of core helium burning
have reached the stellar surface. Since remaining \Al26\ in the stellar core is
efficiently destroyed in helium burning due to neutron captures, standard
stellar evolution models predict that the wind ejection phase of hydrostatically
produced \Al26\ has ended already. 

How much \Al26\ has already decayed? Typical WR lifetimes from observational
constraints and theoretical models are on the order of $\la 5\times 10^5$~yr,
which is shorter than the \Al26\ lifetime by a factor of 2. This could account
for a reduction of observable \Al26\ in the ISM of at most 40\% if the WR star
were close to the end of its evolution. Some models predict WR lifetimes in
excess of a million years for the most massive stars \cite[]{GMeynet:97:al26}
but these models also predict sufficiently large \Al26\ yields such that the
prolonged time available for decay would not reduce the observable amount below
the measured limit. Also note that WR lifetimes strongly depend on the
mass loss description applied in the model.

A straight-forward explanation for the missing 1.8~\MeV\ flux would be sub-solar
metallicity since the \Al26\ yields scale approximately like $Z^2$. 
Yet, sub-solar abundances in the ISM observed in the Vela direction can readily
be understood by dust formation (and therefore depletion of the gaseous phase)
rather than by intrinsic low metallicity \cite[]{EFitzpatrick:96:ism:compos}.
Analysis of spectroscopic data for the O star in the binary system is indeed 
consistent with solar metallicity \cite[]{ODeMarco:99:gamma_vel}.

\begin{figure}
\centering
\psfig{file=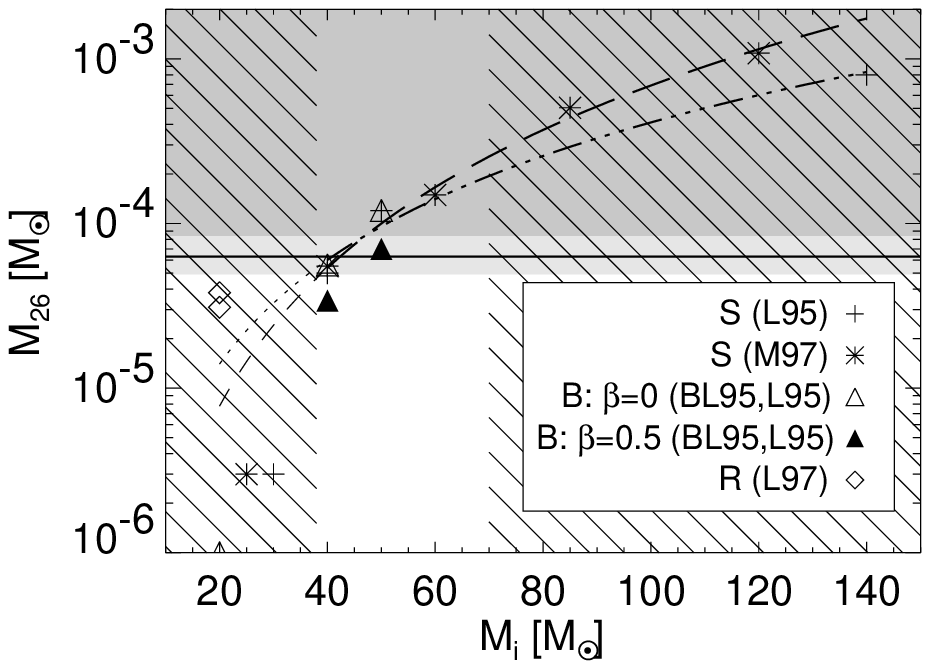,width=\fignarrow}
\caption[]{\Al26\ yields as a function of the initial mass of the WR component
in the binary system. Plotted are single star (``S'') and binary (``B'') models
as well as models of rotating (single) stars (``R''), all at solar
metallicity. L95: \cite{NLanger:95:al26}; M97: \cite{GMeynet:97:al26}; BL95:
\cite{HBraun:95:wr11}; L97: \cite{NLanger:97:rotation}. For interpolation, the
single star models were fitted with a power law (dashed and dash-dotted lines).
The horizontal solid line marks our 2\,$\sigma$ limit (using model c) for the
\Al26\ mass surrounding the WR star. The light-gray region corresponds to the
1\,$\sigma$ distance uncertainty, the dark-gray region is excluded by the
1.8~\MeV\ flux limit. The high-lighted area indicates the most probable initial
mass range. Single star models are barely compatible with our measurement, and
even binary models are consistent only at the lowest allowed initial mass range
and would require significant mass transfer ($\beta$) from the primary to the
secondary.}
\label{f:yields} 
\end{figure}

With these considerations, Fig.~\ref{f:yields} shows that \Al26\ yields
predicted from single star models are barely compatible with the COMPTEL flux
limit. This would suggest that model parameters such as the mass loss
description, which has greatest impact on the stellar structure for initial
masses $\ga 40$~\Msol, or internal mixing parameters like core overshooting or
semi-convection may have to be modified. 

\subsection{Binary models}
Differences in the stellar evolution of the primary star in a relatively wide
binary system such as \gVel\ ($a \approx 1$~AU) stem from mass loss, while
effects of tidal forces on the stellar structure are negligible during the main
sequence phase when \Al26\ is produced in the core. In addition to the mass loss
mechanisms of single stars, Roche Lobe Overflow (RLOF) can change the stellar
structure of both primary and secondary star. For the discussion of \Al26\
yields, we can concentrate on the evolution of the primary since the secondary O
star merely reveals unprocessed material from the stellar envelope at its
surface, but might bury some of the processed material transfered from the
primary. \cite{NLanger:95:review} proposes that binary stars with $M_\mathrm{i}
\ga 40$~\Msol\ undergoing RLOF are essentially indistinguishable from single
stars in the same mass range undergoing a phase of very intense mass loss as
LBVs. Differences in the \Al26\ yield in the circumstellar medium would merely
result from the fraction of expelled mass captured by the secondary companion,
i.e.\ some fraction $\la \beta$ could be buried in the surface layers of the
secondary. (Stellar winds from primary and secondary will always transport some
fraction of \Al26\ into the ISM.) \cite{DVanbeveren:91} argues that this LBV
mass loss may even prevent any occurence of RLOF, which means that \Al26\ yields
should remain the same as for single stars.

For initial masses below $\sim 40$~\Msol, RLOF can provide additional mass
loss not attainable in the single star scenario, pushing down the lower initial
mass limit for the formation of WR stars to about 20~\Msol\
\cite[]{DVanbeveren:98:evolution}. Yet, adopting current masses of 9~\Msol\
for the WR star and 29~\Msol\ for the O star and assuming a fraction of mass 
transfered to the secondary as high as $\beta = 0.5$ would yield a
minimum initial mass for the primary of $\sim 30$~\Msol, given that the primary
star must have been the more massive partner initially to evolve faster. 
The observed current mass loss rate of $(2.8^{+1.2}_{-0.9})\times 10^{-5} \:
\mbox{\Msol/yr}$ \cite[]{DSchaerer:97:wr11}, quite typical for this type of 
star, supports a minimum mass lost into the ISM of at least 10~\Msol\ in
the last few 100,000 years. \cite{DVanbeveren:98:evolution} quote a minimum
initial mass of 38~\Msol\ for the WC star in \gVel\ based on detailed models of
binary evolution.

Overall, the minimum initial mass for the WR star in binary models is found
close to the values obtained by single star models, namely around 40~\Msol.
While the real initial mass may well have been larger, it is apparent from
Fig.~\ref{f:yields} that discrepancies between predicted \Al26\ yields and the
measured 1.8~\MeV\ flux limit become quite severe for larger masses.
Even for $M_\mathrm{i}=40$~\Msol, models are in clearly better agreement with
the flux limit if a significant fraction of the ejected mass of the primary
accreted onto the secondary. The binary models of \cite{HBraun:95:wr11} lead to a
typical reduction in \Al26\ yield of about 40\% for models with $\beta = 0.5$ as
displayed in the figure. 

\begin{figure}
\centering
\psfig{file=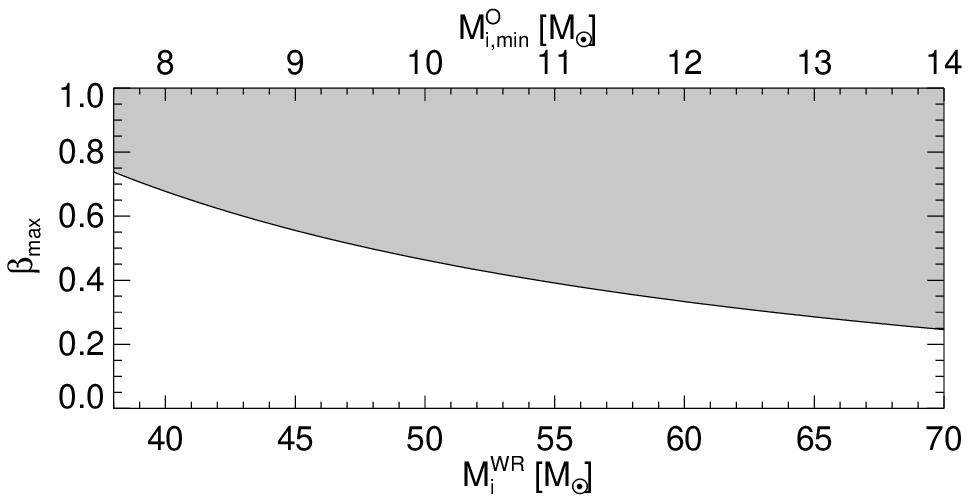,width=\fignarrow}
\caption[]{Maximum value for the mass fraction $\beta$ transfered to the
secondary assuming current masses $M_\mathrm{WR}=9$~\Msol,
$M_\mathrm{O}=29$~\Msol, and an initial mass ratio $q > 0.2$ to avoid merging
\cite[]{DVanbeveren:98:evolution}.} 
\label{f:max_beta} 
\end{figure}  

Information regarding a quantitative estimate of $\beta$ is still sparse, but
some statements can be made. If we consider $M_\mathrm{i} \approx 40$~\Msol\ the
lowest possible initial mass for the WR star, the fact that binaries with
initial mass ratios (secondary / primary) $q < 0.2$ are expected to merge
\cite[]{DVanbeveren:98:evolution} leads to a minimum initial mass for the O~star
of $\sim 8$~\Msol, hence an initial total mass of the system of $\ga
48$~\Msol. Assuming a current mass of $(29+9=38)$~\Msol\ as for the initial mass
estimate, a minimum of 10~\Msol\ must have been expelled to the ISM, while about
20~\Msol\ would have been transfered to the secondary, corresponding to $\beta
\approx 2/3$. Any larger value of $q$, i.e.\ larger initial mass of the O star,
would yield lower $\beta$. For larger values of the initial mass of the WR
star, the maximum allowed $\beta$ for $q>0.2$ rapidly decreases as illustrated
in Fig.~\ref{f:max_beta}.

Observation of an ``ejecta-type'' ring nebula around \gVel\ with 57' angular
diameter \cite[]{AMarston:94:survey2}, corresponding to a radius of $2.1$~pc,
demonstrates that significant mass has been expelled from the system, probably
during a preceding LBV / WN phase, even though the mass of the shell has not yet
been determined. If indeed the IRAS Vela shell were the remnant of the main
sequence bubble of the primary star, the mass transfer to the secondary would be
negligible compared to the ejected mass and the WR star could be reasonably
modeled by a $M_\mathrm{i} \approx 60$~\Msol\ star \cite[]{GGarcia:96:CSM:LBV}.
This scenario is supported by the recent finding that the helium abundance in
the companion O star is not enriched \cite[]{ODeMarco:99:gamma_vel} as one might
expect if significant amounts of processed material had been transfered to the
secondary star. This, however, would imply a major conflict of the \Al26\ yields
predicted by corresponding stellar models with our measurement.
  
There are additional qualitative arguments for both \emph{very little} mass
transfer to the secondary but also \emph{some} mass transfer: The relatively
high excentricity of the orbit favours little mass transfer, which usually tends
to reduce the excentricity. On the other hand, the high rotation velocity of the
secondary of $v_\mathrm{rot} \sin i = 220$~\kms\ \cite[]{DBaade:90} may be the
result of spinning up by accretion. More detailed modeling is required to
resolve these issues.

\section{Conclusion}
Given the small distance of \gVel\ from HIPPARCOS measurements and
predicted \Al26\ yields of current stellar models, the non-detection of
1.8~\MeV\ emission by COMPTEL comes as a surprise. Combined with other
observations regarding current masses, metallicity and mass transfer, only a
very small volume of the model parameter space remains consistent with the
COMPTEL $2\,\sigma$ upper limit of $M_{26}^\mathrm{WR11} <
\left(6.3^{+2.1}_{-1.4}\right) \; 10^{-5}~\Msol$. Single star models are in
conflict with this value. Binary models alleviate the discrepancy
if significant mass transfer to the secondary occured, burying some fraction of
\Al26\ in the surface layers of the O star, and if the initial mass of the WR
star was close to its minimum value of $\sim 40$~\Msol. It may be more likely,
however, that adjustment of some of the model parameters, e.g. the
parametrization of mass loss, is required. 

Unfortunately, \gVel\ is the only known WR star for which there was hope to
obtain a positive detection with current instruments. The WR stars next closest
to the Sun (WR142, WR145, and WR147 in the Cygnus region) are at least a factor
of two more distant, and therefore their expected fluxes are out of reach for
COMPTEL. For the forthcoming INTEGRAL mission, these stars may just become
detectable, and the 1.8~\MeV\ flux from \gVel\ will be tested down to
significantly lower values.  Yet, to probe the 1.8~\MeV\ flux from individual WR
stars within the radius of completeness of the current WR catalogue 
\mbox{($\sim 2.5$ -- 3 kpc)}, would require a next-generation instrument with a
sensitivity of $10^{-7}$~\flux\ and an angular resolution \mbox{$<0.2\deg$}.

\begin{acknowledgements}
The COMPTEL project is supported by the German government through DARA grant
50 QV 90968, by NASA under contract NAS5-26645, and by the Netherlands
Organization for Scientific Research NWO. The authors are grateful for
discussions with Norbert Langer, Orsola De Marco, Georges Meynet, 
Tony Marston, Nikos Prantzos, Daniel Schaerer, and Werner Schmutz. We thank the
referee Dieter Hartmann for helpful comments.
\end{acknowledgements}

\bibliographystyle{aa}
\bibliography{mnemonic_short,myreferences}
\end{document}